\documentclass[english,aps,prl,reprint,superscriptaddress,floatfix,amsfonts]{revtex4-1}
\usepackage[latin9]{inputenc}
\usepackage{color}
\usepackage{bm}
\usepackage{multirow}
\usepackage{amsmath}
\usepackage{graphicx}
\usepackage{times}
\PassOptionsToPackage{normalem}{ulem}
\usepackage{ulem}
\usepackage[dvipsnames]{xcolor}
\makeatletter

\usepackage{epstopdf}
\usepackage{dcolumn}
\usepackage{bm}
\usepackage{babel}
\usepackage{multirow}
\makeatletter

\@ifundefined{textcolor}{}
{%
 \definecolor{BLACK}{gray}{0}
 \definecolor{WHITE}{gray}{1}
 \definecolor{RED}{rgb}{1,0,0}
 \definecolor{GREEN}{rgb}{0,1,0}
 \definecolor{BLUE}{rgb}{0,0,1}
 \definecolor{CYAN}{cmyk}{1,0,0,0}
 \definecolor{MAGENTA}{cmyk}{0,1,0,0}
 \definecolor{YELLOW}{cmyk}{0,0,1,0}
}

\usepackage{babel}

\makeatother

\begin{document}

\title{Enhancement of diamagnetism by momentum-momentum interaction: application to benzene}

\author{Tha\'{i}s V. Trevisan}

\altaffiliation[Present address: ]{Ames Laboratory, Ames, Iowa 50011, USA}
\affiliation{Instituto de F\'isica Gleb Wataghin, Unicamp, Rua  S\'ergio Buarque de
Holanda, 777, CEP 13083-859 Campinas, SP, Brazil}

\author{Gustavo M. Monteiro}

\altaffiliation[Present address: ]{Department of Physics, City College, City University of New York, New York, NY 10031, USA.}

\affiliation{Instituto de F\'isica Gleb Wataghin, Unicamp, Rua S\'ergio Buarque de
Holanda, 777, CEP 13083-859 Campinas, SP, Brazil}

\author{Amir O. Caldeira}

\affiliation{Instituto de F\'isica Gleb Wataghin, Unicamp, Rua S\'ergio Buarque de
Holanda, 777, CEP 13083-859 Campinas, SP, Brazil}

\begin{abstract}
A well-known property of aromatic molecules is their highly anisotropic response to the presence of an external magnetic field: the component of their magnetic susceptibility parallel to the field is generally much larger than the remaining in-plane components. This intriguing phenomenon is rationalized as a consequence of the delocalization of the itinerant electrons that populate the aromatic ring. In this work, we revisit the magnetism of aromatic molecules and propose an extended Hubbard model for the electrons in the aromatic ring that takes into account the interaction between them and the bonding electrons. We show that the bonding electrons play an important and overlooked role: they mediate an effective, attractive momentum-momentum interaction between the itinerant electrons, which promotes a strong enhancement in the magnetic response of the aromatic ring. For the particular case of benzene, we show that the experimentally observed magnetic anisotropy is recovered with realistic values of the coupling constants.
\end{abstract}
\maketitle

An important property of aromatic molecules is the large anisotropy in their magnetic response. As observed for the first time in the 1930s \cite{krishnan1932Nature,Krishnan1933,Pauling1936}, when an aromatic molecule is subjected to an external magnetic field perpendicular to its plane, the component of the induced magnetic moment parallel to the field was found to be much larger than the perpendicular ones. Such anisotropy is reflected in the molecule's magnetic susceptibility tensor, which describes a prolate ellipsoid, with the long axes ($\chi_{\parallel}$) coinciding with the direction of the field. The imbalance between the in-plane and out-of-plane components of the magnetic susceptibility defines the molecular magnetic anisotropy, $\Delta\chi=\chi_{\parallel}-\bar{\chi}_{\perp}$, where $\bar{\chi}_{\perp}$ denotes the average in-plane components \cite{Rolfes1969}.

In the early 1930s, a phenomenological model - today known as \textit{ring current model} (RCM) \cite{Lazzeretti2000} - was developed by L. Pauling \cite{Pauling1936}, F. London \cite{London1937} and K. Londsdale \cite{Lonsdale1937} to explain this curious phenomena. In a nutshell, the RCM model attributes the origin of the large molecular magnetic anisotropy to the itinerant electrons in the aromatic ring: these nearly free electrons move under the influence of the periodic potential generated by the atomic core and bonding electrons. In the presence of an external magnetic field, each of them acquires a momentum component tangential to the ring, resulting in a current loop along the aromatic ring. While the spin degree of freedom and the bonding electrons contributes equally to $\chi_{\parallel}$ and $\chi_{\perp}$, the current loop only contributes to $\chi_{\parallel}$, originating the anisotropy $\Delta \chi$. In other words, the RCM model states that the magnetic anisotropy of the aromatic molecules is a consequence of the orbital degrees of freedom of the itinerant electron along the aromatic ring, and the larger is the size of the aromatic ring, the more pronounced is the phenomenon \cite{Pauling1936}. For benzene, the most celebrated example of aromatic molecule, $\chi_{\parallel}\approx 2.5\bar{\chi}_{\perp}$ \cite{krishnan1932Nature}, leading to a magnetic anisotropy of $\Delta\chi=-5.48\times 10^{-5}\textrm{cm}^{3}/\textrm{mol}$ \cite{Dauben1969,Pauling1936}. 

The simple semi-empirical interpretation of the magnetic properties of the aromatic molecules granted the RCM model with great success. While the search for a precise definition of aromaticity is an ongoing challenge in modern chemistry, several criteria were developed to determine whether a given molecule can be considered aromatic \cite{iupac}. Among these criteria, the most used until today is the \textit{magnetic criteria}, which considers the development of electric currents along a cyclic conjugated atomic structure as a strong evidence of aromaticity \cite{gershoni2015}. Along the years, the original RCM model has been refined by several authors \cite{Lazzeretti2000} to incorporate quantum effects. Curiously, in these refinements, the bonding electrons have always played a less important role: they only contribute to generating the periodic potential, often considered \textit{static}, to which the itinerant electrons are subjected. As a consequence, the effective models available to describe the aromatic molecules take into account only the degrees of freedom of the itinerant electrons in the aromatic ring \cite{wu2002,schuler2013,pariser1953I,pariser1953II,pople1953}.  

In contrast, in this work, we show that the itinerant electrons alone cannot adequately describe the magnetic anisotropy of the aromatic molecules. The bonding electrons play an important, and often overlooked role; although they are localized at the strong electron-pair bonds that keep adjacent atoms together, they undergo virtual excitations triggered by the itinerant electrons themselves. As a result, the effective periodic potential felt by the itinerant electrons  changes dynamically, generating a feedback effect in their dynamics. Formally, in a previous work \cite{TrevisanArxiv} we treated this effect through a perturbative correction to the Bohr-Oppenheimer approximation as applied to the interaction between the itinerant and the bonding electrons. This gives rise to an effective attractive momentum-momentum interaction between the itinerant electrons. Here, we investigate the consequences of such a momentum-momentum interaction to the magnetic properties of aromatic molecules.

We propose a minimal microscopic model for the aromatic molecules consisting of an extended Hubbard model for the itinerant electrons, where the aforementioned momentum-momentum interaction is considered in addition to the standard nearest-neighbor hopping and on-site repulsion. Although our model applies to aromatic molecules in general, in this manuscript we focus on the specific case  of benzene.  We find that the main consequence of the effective momentum-momentum interaction is a strong amplification of the diamagnetic response of benzene. In particular, we recover its experimentally measured magnetic anisotropy for realistic values of the model parameters, which cannot be achieved with the standard Hubbard model alone. This is the central result of this manuscript.  

\begin{figure}[b!]
\includegraphics[width=1\columnwidth]{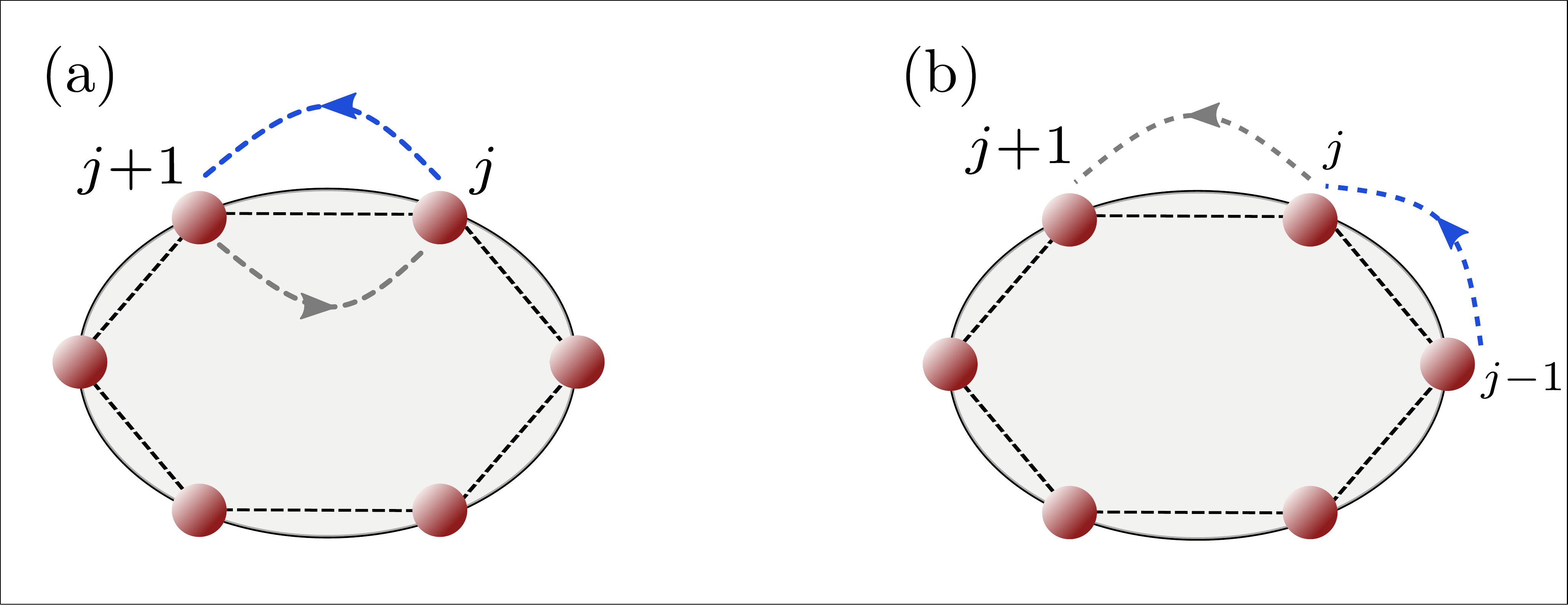}
\caption{Two types of two-body processes that appear in Eq.(\ref{eq:Hpp}). (a) The first term of (\ref{eq:Hpp}) is a bubble-like process that favors the localization of the electrons at the $sp_2$ bonds of the molecule. (b) The second term of (\ref{eq:Hpp}) favors the formation of a loop current along the ring. The Hermitian conjugate of these processes are not represented, as they are trivially obtained by reversing the direction of the arrows.}
\label{Fig:terms}
\end{figure}

Benzene is a planar stable molecule with six carbon atoms arranged in a loop; the \textit{aromatic ring}. In this configuration, the outermost $2s$, $2p_x$, and $2p_y$ carbon orbitals hybridize forming the strong $\sigma$-bonds in the plane of the molecule, which keep the neighbor carbon atoms together, and also bind each of them with a hydrogen atom. The $2p_z$ orbitals, on the other hand, which are perpendicular to the aromatic ring, remain unchanged, and the overlap between neighboring $p_z$ orbitals form weaker $\pi$-bonds. In total, 24 electrons occupy the $\sigma$-bonds, and 6 electrons occupy the $p_z$ orbitals. Hereafter they are denominated $\sigma$-electrons and $\pi$-electrons, respectively. The $\pi$-electrons are delocalized and  referred to as \textit{itinerant} electrons. The $\sigma$-electrons, on the other hand, are localized in the $\sigma$-bonds and are also called \textit{bonding} electrons. We will shortly see that the separation of the energy scales associated with these two types of electrons has important implications for the magnetic properties of the aromatic ring. 

Our microscopic model for benzene consists of an extended Hubbard model $\hat{H}=\hat{H}_0+\hat{H}_{pp}$, whose terms we analyze next. The first contribution, 
\begin{equation}
\hat{H}_{0}=-t\sum\limits_{j=1}^{N}\sum\limits_{\sigma}\left(c_{j\sigma}^{\dag}c_{j+1\sigma}+\text{h.c.}\right)+U\sum\limits_{j=1}^{N}\sum\limits_{\sigma,\sigma'}\hat{n}_{j\uparrow}\hat{n}_{j\downarrow},
\label{eq:H0}
\end{equation}   

\noindent is the standard Hubbard Hamiltonian for the six $\pi$-electrons, where the operator $c_{j\sigma}^{\dag}$ ($c_{j\sigma}^{\null}$) creates (annihilates) an electron with spin $\sigma$ at the $p_z$ orbital of the site $j$. The parameters $t$ and $U$ denote nearest-neighbor hopping and the on-site repulsion, respectively. Nonlocal Coulomb interactions, such as the nearest-neighbor repulsion, are strong in aromatic molecules and, in principle, should be included in Eq.(\ref{eq:H0}). However, it  has been shown that the nearest-neighbor repulsion suppresses  $U$, and an effective single-band Hubbard model of the form of Eq.(\ref{eq:H0}) still holds for the itinerant electrons, as long as $U$ is consistently renormalized \cite{schuler2013}. In particular, for benzene it was estimated in Ref. \cite{schuler2013} a ratio $U/t=1.2$, with $t=2.54 \,\,\textrm{eV}$. We adopt these parameter values throughout this manuscript.  

The second contribution,
\begin{align}
\hat{H}_{pp}=-\lambda\left(\frac{U}{t}\right)^2&\sum\limits_{j=1}^{N}\sum\limits_{\sigma\sigma'}\left[\left(c_{j\sigma}^{\dag}c_{j+1\sigma'}^{\dag}c_{j\sigma'}^{\null}c_{j+1\sigma}^{\null}+\text{h.c.}\right)\right.\nonumber\\
&+\left.\left(c_{j\sigma}^{\dag}c_{j-1\sigma'}^{\dag}c_{j-2\sigma'}^{\null}c_{j-1\sigma}^{\null}+\text{h.c.}\right)\right] 
\label{eq:Hpp}
\end{align}

\noindent accounts for the dynamical changes in the periodic potential felt by the $\pi$-electrons due to the interaction between them and the $\sigma$-electrons. As shown in Ref.\cite{TrevisanArxiv}, the virtual transition of the $\sigma$-electrons mediate an effective interaction between the $\pi$-electrons which, in first quantization, has the form of an attractive momentum-momentum interaction. Note that Eq.(\ref{eq:Hpp}) is similar to the effective inter-electronic interaction mediated by plasmons in an electron-gas derived in the seminal works of Bohm and Pines \cite{bohm1951,bohm1952,bohm1953}. In their case, such effective interaction was negligible due to screening effects. Here, in contrast, the screening effects are not strong enough to suppress $\hat{H}_{pp}$, since we are dealing with a few body-system. In this work, we focus on the half-filling regime ($N=N_e=6$) and keep the ionic cores always static, as our goal is to focus solely on the electronic orbital degrees of freedom. Moreover, since only extremely high temperatures are comparable to the present molecular energy scales, thermal effects play no significant role in the phenomena we address in this manuscript. As a consequence, we consider hereafter the ground state of $\hat{H}$.

In second quantization, the effective momentum-momentum interaction takes the form shown in Eq.(\ref{eq:Hpp}). As illustrated in Fig. \ref{Fig:terms}, it involves two distinct many-body processes: the first term on the right-hand side of Eq.(\ref{eq:Hpp}) is a bubble-like term responsible for the creation and subsequent annihilation of electrons between first-neighbor sites, favoring, therefore, the localization of the $\pi$-electrons around those sites. The second term, on the other hand, plays the leading role in the enhancement of the molecule's diamagnetism, since it favors an ordered motion of the itinerant electrons along the ring.   

\begin{figure}[t!]
\includegraphics[width=0.99\columnwidth]{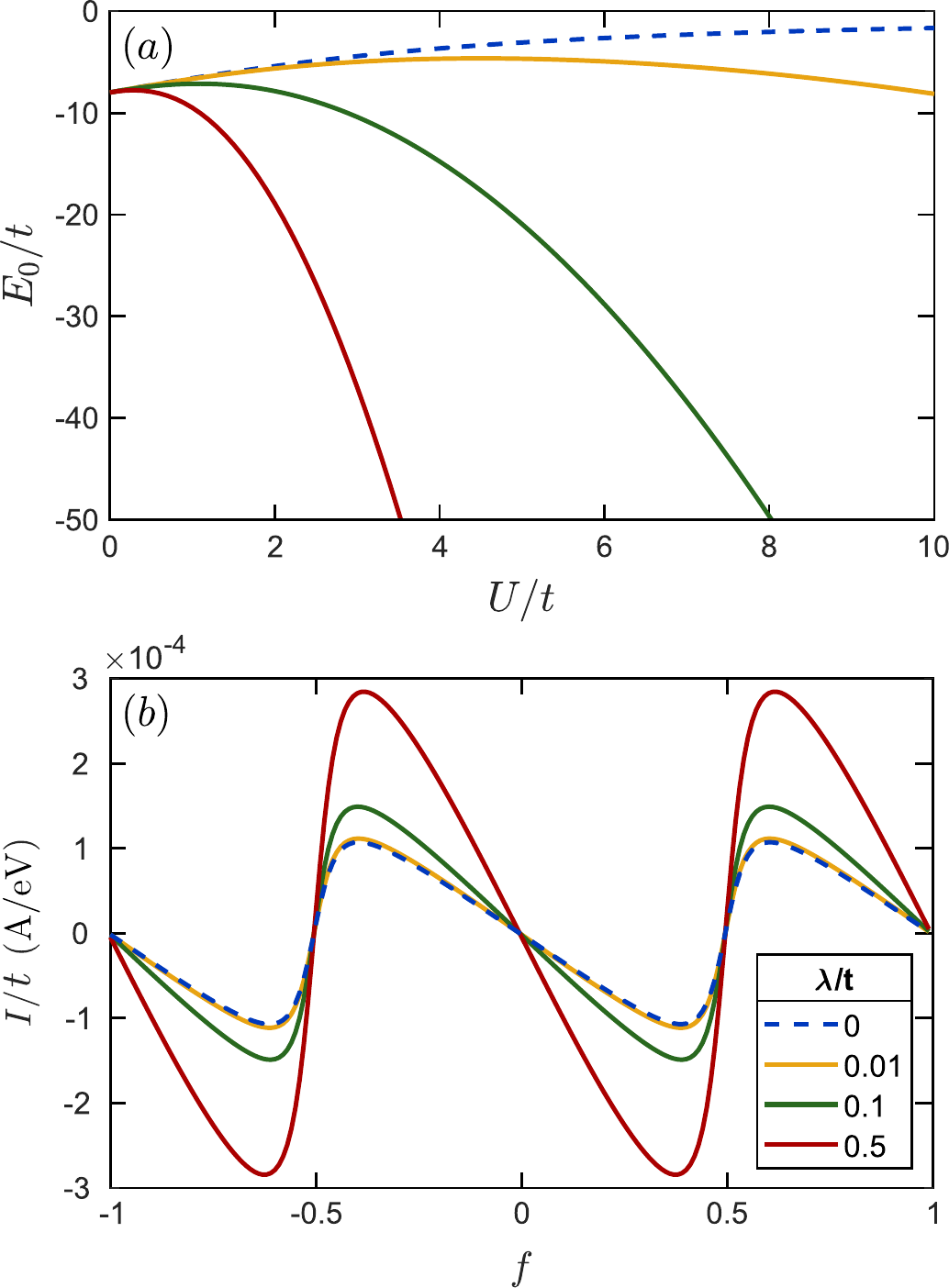}
\caption{(a) Ground state energy of the minimal model for benzene defined in Eqs.(\ref{eq:H0}) and (\ref{eq:Hpp}) as a function of the on-site repulsion for several values of the ratio $\lambda/t$. (b) Persistent current in the aromatic ring when a uniform magnetic field is applied perpendicularly to the plane of the molecule. The current is shown as a function of the magnetic flux that pierces the aromatic ring. The ground state energy, persistent current, and the on-site repulsion are normalized by the hopping amplitude $t$.} 
\label{fig:EandI}
\end{figure} 

The coupling constant $\lambda$ in Eq.(\ref{eq:Hpp}) is given by $\lambda =t^4/\Lambda^3$, where $\Lambda$ denotes the gap of the $\sigma$-electrons to their first excited state \cite{TrevisanArxiv}. Since the $\sigma$-electrons are localized in the bonds, it is more costly to promote them to an excited state than to move the delocalized $\pi$-electrons along the aromatic ring. This is reflected in the relation between $\Lambda$ and $t$, where we have $\Lambda>t$ or, equivalently, $\lambda/t<1$. It is important to note that if $\Lambda\gg t$, the $\sigma$-electrons can be considered frozen in their ground state, since no virtual excitation will be triggered by the itinerant electrons.  The regime we explore here, instead is $\Lambda \gtrsim t$. 

For benzene, we estimate $\Lambda$ by the gap between the highest-energy bonding $\sigma$ molecular orbital (MO) and the lowest-energy anti-bonding $\sigma$-MO. Ionization spectrum measurements \cite{Natalis1968,Brundle1972} reveal that the highest-energy bonding $\sigma$-MO, denoted by $\sigma(e_{2g})$ is in between the two bonding $\pi$-MOs, which are called $\pi(e_{1g})$ and $\pi(a_{2u})$. A simple tight-biding calculation gives energies  $-2t$, $-t$ (doubly degenerate), $t$ (doubly degenerate) and $2t$ for the bonding  $\pi(e_{1g})$ and the $\pi(a_{2u})$, and corresponding anti-bonding $\pi^{*}(e_{1g})$ and the $\pi^{*}(a_{2u})$ $\pi$-MOs, respectively. Note that the energy levels of the anti-bonding $\pi$-MOs are the opposite of the corresponding bonding MOs. The same is true for the $\sigma$-MOs. As a result, the order of magnitude of $\Lambda$ is between $2t$ and $4t$, and, therefore, the effects of the coupling between the bonding and itinerant electrons are not negligible. 

A closer look at Eqs.(\ref{eq:H0}) and (\ref{eq:Hpp}) reveals an interplay between the on-site repulsion and the momentum-momentum interaction. While the on-site repulsion depends linearly on $U/t$, the magnitude of the momentum-momentum interaction grows with $(U/t)^2$ and its effects dominates over the on-site repulsion as $U$ increases. Such interplay becomes evident in the non-monotonic behavior of the ground state energy $E_0$ as a function of the on-site repulsion, which is obtained through the exact diagonalization of $\hat{H}$. As shown in Fig.\ref{fig:EandI}(a), $E_{0}$ initially increases with $U$, since the on-site repulsion tends to localize the electrons at the ring's sites. However, the momentum-momentum interaction becomes more attractive and, for large enough $U$, leads to a downturn in the ground state energy. Consistently, the larger is $\lambda/t$, the sooner (smaller values of $U/t$) this downturn takes place. We emphasize that the trend of $E_0$ as a function of $U$ results from an energy competition between the different terms of the Hamiltonian instead of a competition between qualitatively different ground states.  

Our goal is to study the consequences of the effective momentum-momentum interaction (\ref{eq:Hpp}) to the magnetic properties of benzene. For this purposes, we apply a uniform magnetic field oriented perpendicularly to the molecular plane. In the presence of this field, Eqs.(\ref{eq:H0}) and (\ref{eq:Hpp}) needs to the slightly modified, as the creation and annihilation operators acquire a complex phase proportional to the magnetic flux $\phi$ enclosed by the aromatic ring. Accordingly, for the hopping term $c_{j\sigma}^{\dag}c_{j+1\sigma}^{\null}\rightarrow e^{i 2\pi f/N} c_{j\sigma}^{\dag}c_{j+1\sigma}^{\null}$, while the on-site repulsion remains unchanged. Moreover, the momentum-momentum interaction becomes
\begin{align}
\hat{H}_{pp}^{(mag)}&=-\lambda\left(\frac{U}{t}\right)^2\sum\limits_{j=1}^{N}\sum\limits_{\sigma\sigma'}\left[\left(c_{j\sigma}^{\dag}c_{j+1\sigma'}^{\dag}c_{j\sigma'}^{\null}c_{j+1\sigma}^{\null}+\text{h.c.}\right)\right.\nonumber\\
&+\left.\left(e^{-i4\pi f/N}c_{j\sigma}^{\dag}c_{j-1\sigma'}^{\dag}c_{j-2\sigma'}^{\null}c_{j-1\sigma}^{\null}+\text{h.c.}\right)\right]\text{ .}
\label{eq:newHpp}
\end{align}

\noindent Here, $f=\phi/\phi_0$ denotes the dimensionless magnetic flux, where $\phi_0=h/e$ is flux quanta. The Zeeman splitting was not included above because we focus solely on the magnetic properties of benzene due to the orbital degrees of freedom of its $\pi$-electrons. The spin degrees of freedom generate an isotropic magnetic response and, therefore, do not contribute to the molecule's magnetic anisotropy. 

The magnetic field induces an angular momentum component to each of the $\pi$-electrons, and an electric current  flows around the ring. It is given by
\begin{equation}
I(f)=-\frac{1}{\phi_0}\frac{\partial E_0(f)}{\partial f}\text{,}
\label{eq:current}
\end{equation}
\noindent.

\noindent Here, $E_{0}(f)$ is the ground state energy, an oscillatory function of the magnetic flux $f$, given the complex phases acquired by the Hamiltonian [see Eq.(\ref{eq:newHpp})]. Consistently, with the downturn of $E_0$, the momentum-momentum interaction promotes an enhancement of  the ground-state current, as evidenced in Fig. \ref{fig:EandI}(b). Importantly,  Eq.(\ref{eq:current}) corresponds to a \textit{persistent current}, since it does not suffer effects of dissipation. However, it should not be confused with a supercurrent that develops in a superconductor loop since their nature are completely different. While the supercurrent is a result of the condensation of a macroscopic number of Cooper pairs, here the absence of dissipation is a consequence of quantum coherence of the $\pi-$electrons moving along the ring, and it vanishes as soon as the magnetic field is turned off \cite{imry2002,bouchiat2008}. In other words, Eq.(\ref{eq:current}) is a \textit{normal state persistent current}.

\begin{figure}[t!]
\includegraphics[width=0.95\columnwidth]{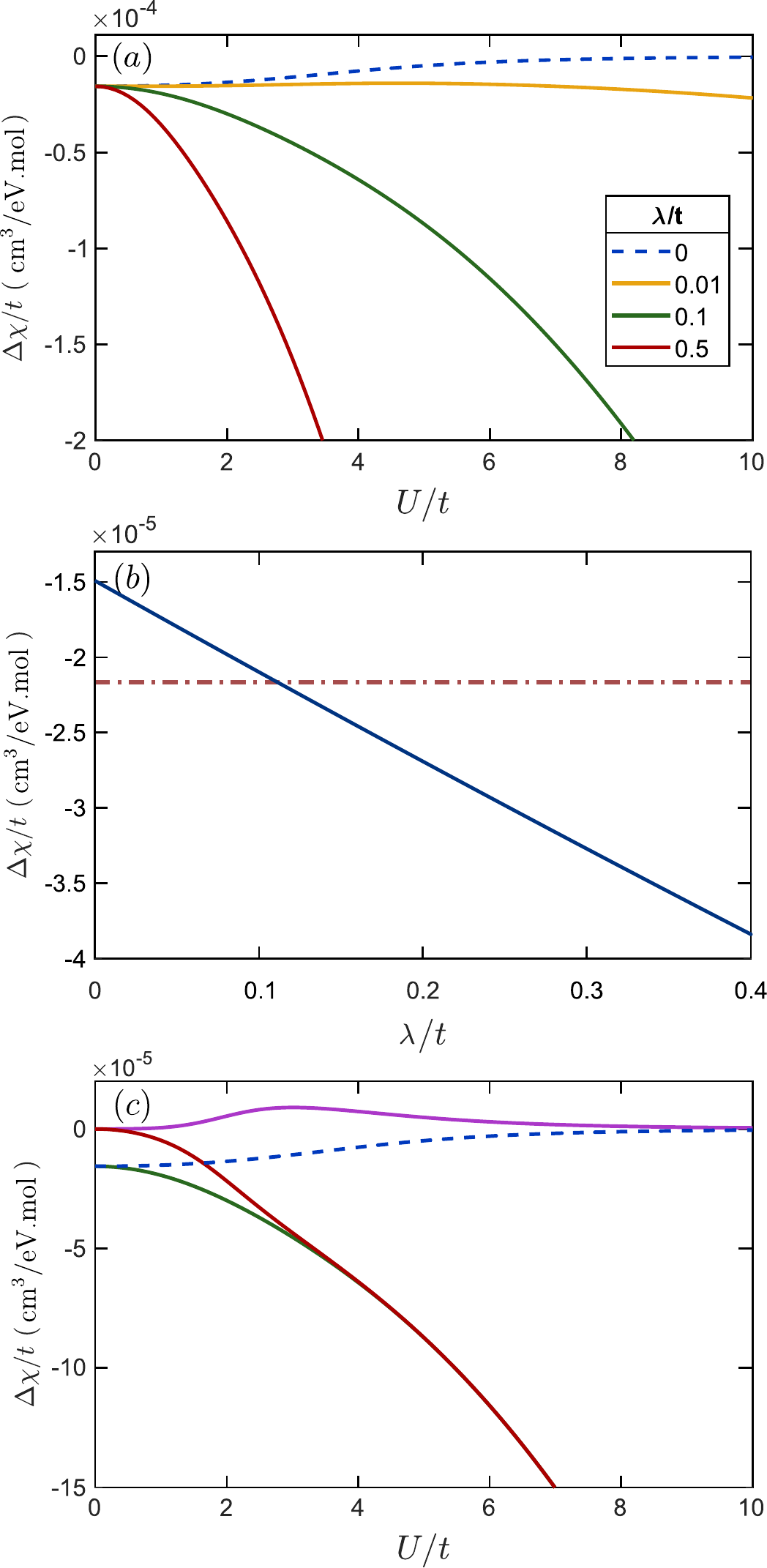}
\caption{(a) Magnetic anisotropy normalized by the hopping amplitude as a function of the on-site repulsion for several values of the ratio $\lambda/t$. Note that $\lambda=0$ corresponds to the standard Hubbard model [Eq.(\ref{eq:H0})]. (b) (Solid blue line) Magnetic anisotropy as a function of $\lambda/t$ with $t=2.54\,\textrm{eV}$ and $U/t=1.2$. The dashed red line shows the experimental magnetic anisotropy for benzene \cite{Dauben1969} normalized by $t$. (c) Different contributions for the magnetic anisotropy. As in panel (a), the dashed blue curve shows the magnetic anisotropy due to the standard Hubbard model alone, while the solid green curve shows the total anisotropy for $\lambda/t=0.1$. The solid purple (red) curve singles out the contribution of the bubble (delocalized) term in Eq.(\ref{eq:Hpp}) to the magnetic anisotropy.}
\label{fig:suscep}
\end{figure}

We saw that a uniform magnetic field, when applied perpendicularly to the molecular plane, generates a non-dissipative current loop along the ring. Such a loop gives rise to a magnetic moment which, by symmetry, is parallel to the external field. As a consequence, the magnetic response of the $\pi$-electrons is purely parallel to the field, and the magnetic susceptibility obtained through the derivative of Eq.(\ref{eq:current}) with respect to $f$,    
\begin{equation}
\Delta \chi=-\frac{N_A(Na)^4\mu_0}{16\pi^2\phi_0^2} \left.\frac{\partial^2 E_0(f)}{\partial f^2}\right|_{f=0} \text{ .}
\label{eq:magsuscep}
\end{equation}

\noindent give us the \textit{magnetic anisotropy} of the ring. Here $N=6$ is the total number of sites of the aromatic ring, $a$ is the lattice spacing  ($a=1.4\,\textrm{\AA}$ for benzene), and $\mu_0$ is the vacuum permeability. Besides, $N_A$ is the Avogadro number, so Eq.(\ref{eq:magsuscep}) expresses the molar magnetic susceptibility. 

Setting $\lambda=0$, Eq.(\ref{eq:magsuscep}) gives the magnetic anisotropy of benzene due to the $\pi$-electrons \textit{alone}, since in this case the $\sigma$-electrons are frozen in the bonds and the low-energy physics of the molecule is described by the standard Hubbard model [Eq.(\ref{eq:H0})]. The dashed blue curve in Fig. \ref{fig:suscep}(a) shows such response. Note that $\Delta\chi$ is suppressed by the on-site repulsion because the larger $U$ is the stronger is the tendency of localization of  the $\pi$-electrons, until no current flows along the ring and $\Delta\chi\rightarrow 0$. Recall that for benzene \cite{schuler2013} $U/t=1.2\,\textrm{eV}$ and $t=2.54\,\textrm{eV}$. For these values of 
the parameters, Eq.(\ref{eq:magsuscep}) gives $\Delta\chi=-3.78 \cdot 10^{-5}\,\textrm{cm}^3/\textrm{mol}$, roughly $3/5$ of the experimental value $-5.48\cdot 10^{-5}\,\textrm{cm}^3/\textrm{mol}$. Not even at $U=0$ the magnetic anisotropy obtained with the standard Hubbard model matches the experimentally observed value, and it would be necessary a hopping parameter roughly twice as big to recover the experimental result.  

The scenario significantly changes once we introduce the momentum-momentum interaction, which strongly enhances $\Delta\chi$ in comparison with the standard Hubbard model alone. Such an effect could have been anticipated from Fig. \ref{fig:EandI}(b), as the momentum-momentum interaction not only increases the magnitude of the persistent current but also makes it steeper at low fields. Indeed, the solid lines in Fig.\ref{fig:suscep}(a) shows that the magnetic response becomes more diamagnetic as $U$ increases when $\lambda\neq 0$. The larger $\lambda$ is, the stronger is the enhancement of the diamagnetic response of the aromatic ring, as shown in Fig.\ref{fig:suscep}(b). In particular, using the values of $U$ and $t$ for benzene \cite{schuler2013}, the susceptibility calculated through Eq.(\ref{eq:magsuscep}) coincides with the experimental value for $\lambda/t\approx0.11$, which corresponds to $\Lambda/t\approx 2.1$. This result is in agreement with our earlier assumption that $t$ and $\Lambda$ were of the same order.

From the two terms in the momentum-momentum interaction [Eq.(\ref{eq:Hpp})], it is the one involving the consecutive hopping between two next-neighbor sites [see Fig.\ref{Fig:terms}(b)] that contributes for the enhancement of the magnetic anisotropy of benzene, since it favors the delocalization of the $\pi$-electrons. This result is highlighted in Fig.\ref{fig:suscep}(c), where we show, separately, the contributions of the standard Hubbard Hamiltonian [Eq.(\ref{eq:H0})], the bubble term of Eq.(\ref{eq:Hpp}), and the delocalized term of Eq.(\ref{eq:Hpp}) to the magnetic anisotropy. While the bubble term gives an almost vanishing contribution for $\Delta\chi$ over the entire range of $U$, the delocalized term gives a strong diamagnetic response and dominates the behavior of the susceptibility as $U$ increases.  
 
\textit{Conclusions-} We highlight the importance, often overlooked, of the interaction between the bonding and the itinerant electrons to the magnetic properties of aromatic molecules. In contrast to the majority of lattice models for aromatic molecules, we relaxed the often adopted constraint that the bonding electrons are frozen in the chemical bonds. In this case, virtual excitations of the bonding electrons, which are triggered by their interaction with the itinerant electrons, mediates an attractive momentum-momentum interaction between the mobile $\pi$-electrons. Here, we show that such an interaction competes with the Hubbard on-site repulsion and favors the electronic delocalization, leading, in the presence of a uniform magnetic field, not only to the development of more intense persistent currents in the system's ground state but, more importantly, to a strong amplification of the molecule's magnetic response in the same direction of the applied field. For the specific case of benzene, we observed a strong enhancement of the diamagnetic susceptibility, recovering the experimental value for $\lambda/t\approx0.11$. Notice that if a much larger value of this coupling constant were necessary to recover the experimental anisotropy, it would imply that $\Lambda\gg t$ and the inclusion of the momentum-momentum interaction in the effective Hamiltonian of the $\pi$-electrons would not be justifiable in the first place. In this case, one could argue that the magnetic anisotropy of benzene would not be affected by the interaction between the $\sigma$- and $\pi$-electrons. Here, however, we obtain $\Lambda/t\approx 2.1$, which is not so far from an estimate of the $\sigma$-electrons gap obtained through benzene ionization spectra. 

\begin{acknowledgments}
We thank Rene Nome for fruitful discussions. We acknowledge S\~{a}o Paulo Research Foundation (FAPESP) and Conselho Nacional de Desenvolvimento Cient\'ifico e Tecnol\'ogico (CNPq) for the financial support. TVT and GMM were supported by FAPESP under the fellowships 2015/21349-7 and 2016/13517-0, respectively. AOC was supported by CNPq under the grant 302420/2015-0.
\end{acknowledgments}

\bibliography{references} 

\end{document}